# Recalculating the Orbit of α-Centauri AB

*Suryadi Siregar* and *Hanindyo Kuncarayakti*
Astronomy Research Division and Bosscha Observatory,
Faculty of Mathematics and Natural Sciences, Bandung Institute of Technology, Bandung
e-mail: suryadi@as.itb.ac.id

*Abstract*

*The two main components of the closest star system, α Centauri AB (RA $14^h39^m$, Dec $-60^o50'$, J2000.0) is indubitably one of the most studied visual double stars. This paper presents the results of our recalculation of orbital and physical parameters of the system using Thiele–van den Bos method, based on observational data from year 1900 to 2002. Despite some significant discrepancies, in general our results confirmed previous results of orbital parameter determinations using different method.*

*Abstrak*

*Dua komponen utama sistem bintang terdekat α Centauri AB (RA $14^h39^m$, Dec $-60^o50'$, J2000.0) merupakan salah satu pasangan bintang ganda yang paling sering dipelajari. Makalah ini menunjukkan hasil penghitungan ulang parameter orbit dan parameter fisik sistem bintang ganda ini dengan menggunakan metode Thiele-van den Bos, menggunakan data observasi dari tahun 1900 sampai tahun 2002. Meskipun terdapat beberapa simpangan yang signifikan, secara umum hasil kami sesuai dengan hasil-hasil penentuan parameter orbital sebelumnya yang dikerjakan dengan metode berbeda.*

*Kata kunci*: Bintang ganda visual- Penghitungan parameter orbit

## 1. Introduction

Visual double stars are very valuable for astrophysics. Accurate determinations of orbital and physical parameters of double star systems give the mass-luminosity ratio for the stars. With distance of about 4.4 light years, α Centauri is the closest star system to the Sun, and has been extensively studied either as a double star system or as individual star. The system consists of three stars; two of them are a well-studied visual double star pair.

According to Washington Double Star Catalogue (WDS-http://ad.usno.navy.mil/wds), the binary nature of α Centauri AB = LDS 494 = WDS 14396-6050 was discovered by Richaud in December 1689, hence its discoverer-code name is RHD 1 AB. The third or C component of this system is located about 9000 arc seconds away from the AB pair; it is the flare star V645 Cen and is known as Proxima Centauri. The first measurement of separation ($\rho$) and position angle ($\theta$) was reported by Lacaille in 1752.

Since Lacaille's measurements, there were many attempts to observe this pair. The Washington Double Star Catalogue recorded over 400 data points of $\rho$ and $\theta$ between epoch 1752.20 and 2002.688 for the pair (Figure 1). The last report of orbital and physical parameters was given by Pourbaix *et al.* (2002). In this paper we present our results in determining the orbital and physical parameters, and the comparison with other determinations.

This work is a part of visual double star research carried out at the Bosscha Observatory since its establishment in 1923. Between 1950 and 2000 the photographic observations of visual double stars played an important role in our observatory. In the last decade some observational data of photographic plates of α Centauri AB were measured by Jasinta and Soegiartini (1994).

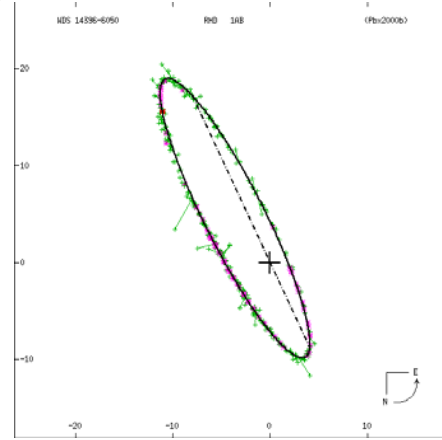

**Figure 1**. The orbit diagram of α Centauri AB from WDS Catalogue. Plus sign is the position of central star. Dashed line is the major axis. North is down and east is to the right.

## 2. Data Acquisition and Pre-process

We used the Washington Double Star Catalogue as the source of data in our calculation. For quality reason, we rejected data older than year 1900 and those showing discrepancies from the $\rho_t$ and $\theta_t$ curves. With this process 250 data points remained, i.e. the data from 1900 to 2002. Figure 2 shows one orbital period covered by the data.

The next step is to correct the data from lunisolar precession effect, $\Delta\theta = 0.00557 \sin (RA) \sec (Dec) (2000-t)$, where $t$ is the epoch of the datum (Siregar, 2006). We employed linear interpolation method to produce a dataset of $\rho_t$ and $\theta_t$ in a four-year interval, resulting in 26 data points between epoch 1902.0 and 2002.0.

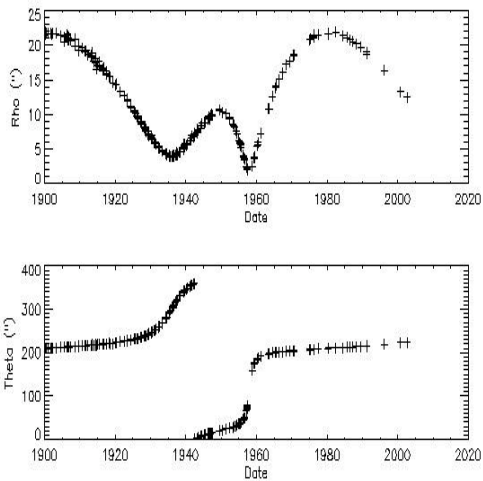

**Figure 2**. Variation of $\rho$ (upper) and $\theta$ (lower) through each epoch.

## 3. Orbital Elements

For the determination of orbital solution we use the Thiele–van den Bos method which is an elaboration of Kepler's second law. This law states that areas swept by the radius vector within equal time interval are equal. Since we have resampled our data into uniform time interval, we expect the area between two data points would all be equal. These areas were represented by Kepler's constant,

$$C = \rho_t \rho_{t+\Delta t} \frac{d\theta}{dt}. \qquad (1)$$

Contrary to our expectations, we found that there were significant discrepancies among data points (Figure 3), which means that they do not obey the law of equal areas. In order to proceed with the calculations using this method, we have to rerrange the $(\rho, \theta)$ data into a data set which obey the law of equal areas.

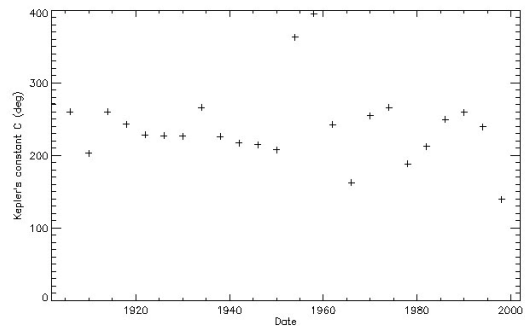

**Figure 3**. Variations of $C$ through each epoch. Average of $C = 241.174$ deg arcsec$^2$ yr$^{-1}$, with standard deviation, $\Delta C = 52.794$ deg arcsec$^2$ yr$^{-1}$.

During the process of rearranging, we used the original $(\rho, \theta)$ dataset as the guideline to prevent excessive correction to the data (Figure 4). To further ensure the accuracy of this process, we also used the orbit diagram as a guideline (Figure 5). The orbit diagram was constructed using equations

$$x = \rho \sin \theta,$$
$$y = \rho \cos \theta. \qquad (2)$$

The orbit diagram consists of data points representing position of the secondary component on celestial plane, relative to the primary. Since those data points are of equal time interval, the distribution of data points along the orbit in orbit diagram indicates the orbital velocity. The crowding effect appearing in the lower left portion of orbit around $(x,y) = (-10,-18)$ indicates positions where the orbital velocity is low, i.e. the rough position of the apastron, while the periastron is roughly around $(x,y) = (0,3)$.

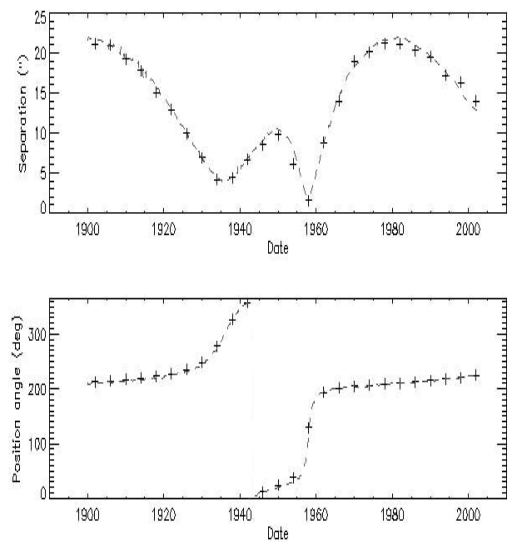

**Figure 4**. Variation of modified $\rho$, and $\theta$ (crosses) compared to original data (dashed line).

The step of rearranging yields the average of $C=222.664$ deg arcsec$^2$ yr$^{-1}$ and standard deviation of $\Delta C=0.036$ deg arcsec$^2$ yr$^{-1}$ (Figure 6). After obtaining the Kepler's constant, then we selected three points which represent the first ($t_1$), middle ($t_2$), and last ($t_3$) data points, which are for the years 1902, 1952, and 2002, respectively. Using these data we then calculated the difference between eccentric anomalies $E_{21}$, $E_{32}$, $E_{31}$ and annual angular speed $\mu$ using the equation

$$\left(t_{qp} - \frac{\Delta_{qp}}{C}\right)\mu = E_{qp} - \sin E_{qp}, \tag{3}$$

where $\Delta_{qp}$ represents areas between two consecutive epochs. This is Thiele's basic equation, which is actually Kepler's equation in different form. This kind of equation is not to be solved analytically, but using iterative numerical techniques. We employed Newton-Raphson method for this purpose.

From the three data points, we solve three similar equation simultaneously to obtain the optimum $\mu$ for which $E_{31}-E_{32} = E_{21}$. We obtained $\mu = 4.4769$, corresponding to orbital period $P=80.4127$ years through the relationship $P = \frac{2\pi}{\mu}$.

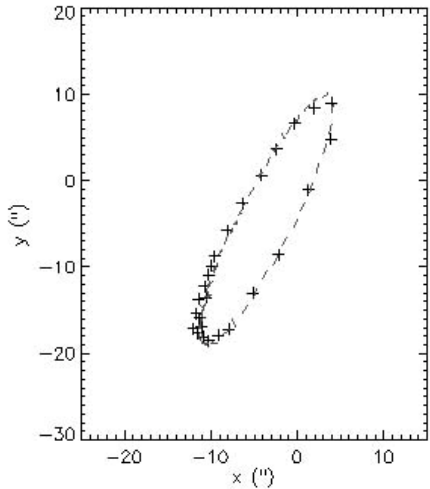

**Figure 5**. Orbit diagram of modified data (crosses) compared to original data (dashed line). North is up and east is to the left.

Using equations in Siregar (2006), we determined the eccentricity to be $e = 0.5147$. Epoch of periastron passage was determined to be $T = 1954.97$, from Kepler's equation $M = E - e \sin E = \mu(t-T)$.

For all data points we calculated the mean anomaly $M$, eccentric anomaly $E$, and parameters $X$ and $Y$. This step enabled us to determine the four Thiele-Innes constants $A, B, F, G$. These constants then were used to calculate the rest of the orbital elements (Siregar, 2006), we obtained:

$A = 9.6972 \pm 0.2093$,
$B = 7.6741 \pm 0.3714$,
$F = -9.8467 \pm 2.2174$,
$G = 15.8912 \pm 3.9347$.

Using the four Thiele-Innes constants, we calculated the longitude of periastron $\omega = 195.07°$, position angle of the nodal line $\Omega = 199.329°$, orbital inclination $I = 73.068°$, and semi-major axis of the orbit $a = 15.49"$.

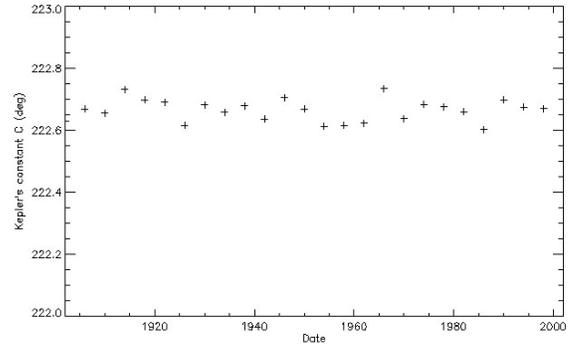

**Figure 6**. Variation of Kepler's constant according to epoch of observations

### 4. Distance and Mass

To determine the distance to the system and masses of each component of $\alpha$ Centauri AB, we employed the dynamical parallax method, given by the equation

$$p = \frac{a}{\sqrt[3]{P^2(M_A + M_B)}}. \tag{4}$$

Here the goal is to determine parallax $p$ and masses $(M_A, M_B)$ by iterations, while keeping $a$ and $P$ constant. We used the values of $a$ and $P$ from our previous determinations, initial value of $p=1.0$ pc, and $(M_A, M_B)$ derived from the absolute bolometric magnitudes using mass-luminosity relationship (Siregar 2006):

$$\log M = 0.1 (4.6 - M_{bol}). \tag{5}$$

The absolute bolometric magnitudes were derived from distance modulus, $M_{bol} = m_{bol} + 5 + 5\log p$. Here we adopt apparent magnitudes of the primary and secondary as 0.14 and 1.24 mag (WDS Catalogue), and bolometric corrections of -0.20 and -0.35, respectively (Drilling and Landolt, 2000).

After seven iterations, the program converged to dynamical parallax, $p = 0".686$ corresponding to the distance $d = 1.458$ pc or equivalent to 4.752 ly. The derived masses for the two components were $M_A = 1.018$ and $M_B = 0.764$ in solar units.

## 5. Ephemeris

By using the orbital elements derived previously, we were able to calculate ephemeris for α Centauri AB. The result is shown in Table 1.

**Table 1**. Ephemeris for α Centauri AB

| Epoch | ρ (″) | θ(deg) |
|---|---|---|
| 2006 | 19.903 | 254.506 |
| 2008 | 19.817 | 260.653 |
| 2010 | 19.543 | 266.616 |
| 2012 | 19.060 | 272.427 |
| 2014 | 18.360 | 278.140 |
| 2016 | 16.654 | 288.075 |

## 6. Discussion

It is interesting to compare this work with previous studies, which used various methods. As an example, in this paper we compare our results with that of Pourbaix *et al.* (2002) which measured the position of secondary relative to primary and radial velocities. Their data consisted of 37 data of primary and 44 data of secondary. Their set of orbital elements and standard error and our calculation are compared in Table 2. Seven out of ten from our derived parameters are within 10% discrepancy compared to theirs. The source of this discrepancy quite probably is the inaccurate data rearranging process during the determination of the Kepler's constant.

The derived masses for both primary and secondary components are in concordance with spectral type G2V and K1V, respectively (Drilling and Landolt, 2000), although different significantly with Pourbaix et al. (2002) for the secondary mass. This result is in agreement with the spectral types reported in WDS. Further efforts in determining orbital solutions (e.g. using different methods) for the pair would be beneficial.

**Table 2**. Orbital and physical parameters of *α* Centauri AB

|  | Pourbaix et al. (2002) | This work | Discrepancy |
|---|---|---|---|
| $P$ (yr) | 79.91±0.011 | 80.41 | 0.63% |
| $a$ (″) | 17.57±0.022 | 15.49 | 11.84% |
| $i$ (°) | 79.20±0.041 | 73.068 | 7.75% |
| $\Omega$ (°) | 204.85±0.084 | 199.32 | 2.69% |
| $T$ (yr) | 1955.57±0.012 | 1954.97 | 0.03% |
| $E$ | 0.5179±0.00076 | 0.5147 | 0.32% |
| $\omega$ (°) | 231.65±0.076 | 195.07 | 15.79% |
| $p$ (″) | 0.747±0.0012 | 0.686 | 8.17% |
| $M_A$ ($M_\odot$) | 1.105±0.0070 | 1.018 | 7.87% |
| $M_B$ ($M_\odot$) | 0.934±0.0061 | 0.764 | 18.20% |

Stellar masses in second column of Table 2 are calculated by using the trigonometry parallax of Soderhjelm (1999). In the third column standard errors are not given because the orbital elements were calculated only for one set of data $\rho$ and $\theta$. The error propagation during iteration is limited up to ε~0.0001. Other information about the age of the system depends critically on whether or not α Centauri AB has a convective core. If it does, then the age of α Centauri AB is ~7.6 Gyr, otherwise this binary system has an age of ~6.8 Gyr (Guenther and Demarque, 2000). Meanwhile earlier study by Demarque *et al.* (1986) gave an age of ~ 4-4.5 Gyr.

## Acknowledgements

We would like to thank Dr. B. Dermawan for his supports and valuable suggestions. We would also like to acknowledge P. Irawati and N. Hasanah for fruitful discussions. This research has made use of the Washington Double Star Catalog maintained at the U.S. Naval Observatory.